\documentclass[12pt]{article}

\usepackage{epsfig}
\usepackage{cite}
\usepackage{amsmath, amssymb, amsfonts}
\usepackage{color}
\usepackage{latexsym}
\usepackage{graphicx}
\usepackage[colorlinks,bookmarks]{hyperref}
\hypersetup{pdfpagemode=UseNone, pdfstartview=FitH, linkcolor=blue,
            citecolor=red, urlcolor=blue}

\def\be{\begin{equation}}
\def\ee{\end{equation}}
\def\ba{\begin{eqnarray}}
\def\ea{\end{eqnarray}}

\def\bdm{\begin{displaymath}}
\def\edm{\end{displaymath}}
\def\la{~\mbox{\raisebox{-.6ex}{$\stackrel{<}{\sim}$}}~}

\def\bq{\begin{quote}}
\def\eq{\end{quote}}

 at 10truept
 at 14truept

\newcommand{\p}{\partial}

\newcommand{\Mpl}{M_{\mathrm{Pl}}}
\newcommand{\mps}{M_{\mathrm{Pl}}^2}

\newcommand{\bea}{\begin{eqnarray}}
\newcommand{\eea}{\end{eqnarray}}

\newcommand{\bi}{\begin{itemize}}
\newcommand{\ei}{\end{itemize}}

\newcommand{\beq}{\begin{equation}}
\newcommand{\eeq}{\end{equation}}
\newcommand{\beqa}{\begin{eqnarray}}
\newcommand{\eeqa}{\end{eqnarray}}
\newcommand{\mpl}{\Mpl}

\def\la{~\mbox{\raisebox{-.6ex}{$\stackrel{<}{\sim}$}}~}

\def\ltap{\ \raise.3ex\hbox{$<$\kern-.75em\lower1ex\hbox{$\sim$}}\ }
\def\gtap{\ \raise.3ex\hbox{$>$\kern-.75em\lower1ex\hbox{$\sim$}}\ }
\def\gl{\ \raise.5ex\hbox{$>$}\kern-.8em\lower.5ex\hbox{$<$}\ }
\def\roughly#1{\raise.3ex\hbox{$#1$\kern-.75em\lower1ex\hbox{$\sim$}}}

\begin{document}

\thispagestyle{empty}
\begin{flushright}
April 2022 \\
DESY 22-070
\end{flushright}
\vspace*{1.5cm}
\begin{center}

{\Large \bf A Quantum-Mechanical Mechanism for Reducing}
\vskip.3cm
{\Large \bf the Cosmological Constant}  

\vspace*{1.15cm} {\large 
Nemanja Kaloper$^{a, }$\footnote{\tt
kaloper@physics.ucdavis.edu} and 
Alexander Westphal$^{b,}$\footnote{\tt alexander.westphal@desy.de}}\\
\vspace{.2cm} 
{\em $^a$QMAP, Department of Physics and Astronomy, University of
California}\\
\vspace{.05cm}{\em Davis, CA 95616, USA}\\
\vspace{.2cm} $^b${\em Deutsches Elektronen-Synchrotron DESY, Notkestr. 85, 22607 Hamburg, Germany}\\

\vspace{1.65cm} ABSTRACT
\end{center}
We exhibit a mechanism which dynamically adjusts cosmological constant toward $0^+$. 
The adjustment is quantum-mechanical, discharging cosmological constant in
random discrete steps. It renders de Sitter space unstable, and triggers its decay 
toward Minkowski. Since the instability dynamically stops at $\Lambda = 0$, the 
evolution favors the terminal Minkowski space without a need for anthropics. The 
mechanism works for any QFT coupled to gravity.

\vfill \setcounter{page}{0} \setcounter{footnote}{0}

\vspace{1cm}
\newpage

\vspace{1cm}

Following the discussion \cite{Kaloper:2022oqv,Kaloper:2022utc} generalizing 
General Relativity (GR) \cite{Hilbert:1915tx,Einstein:1915ca} to a theory of gravity on 
the multiverse, in this {\it Letter} we zoom in on the cosmological constant adjustment to zero.
We focus on a greatly simplified limit of the theory of \cite{Kaloper:2022oqv,Kaloper:2022utc},
with fixed Planck scale\footnote{E.g. by decoupling the membranes of  \cite{Kaloper:2022oqv,Kaloper:2022utc} 
that might change it.}. We allow dynamical variation of only the cosmological constant counterterm, mediated 
by a system of $4$-forms and their membrane sources. This theory generalizes unimodular formulation of GR 
\cite{Einstein:1919gv,Anderson:1971pn,Aurilia:1980xj,Duff:1980qv,Buchmuller:1988wx,Buchmuller:1988yn,Henneaux:1989zc,Ng:1990xz,Buchmuller:2022msj},
by including charged membranes, and corresponding boundary terms which enforce
local general covariance. Since the Planck scale is fixed, we can couple any quantum field theory (QFT) of matter to 
gravity minimally, as there is no chance for ghosts to arise in this limit \cite{Kaloper:2022oqv,Kaloper:2022utc}. 
This suffices to ensure de Sitter space is unstable to membrane nucleations, which 
completely cancel cosmological constant to $0^+$. The huge numerical disparity between the QFT cutoff 
and the observation is irrelevant. It decays away. 

This is how it works. {\it Sans} membranes, cosmological constant is fixed, albeit completely arbitrary. 
It is not correlated to the local QFT scales in a calculable way
\cite{Einstein:1919gv,Anderson:1971pn,Aurilia:1980xj,Duff:1980qv,Buchmuller:1988wx,Buchmuller:1988yn,Henneaux:1989zc,Ng:1990xz,Buchmuller:2022msj};
it is set by initial conditions. However, with membranes, which source $4$-forms, and by extension the cosmological 
constant, quantum discharge changes the {\it physical}  
cosmological constant $\Lambda$ \cite{Brown:1987dd,Brown:1988kg}. 
In a nested set of expanding bubbles bounded by membranes, $\Lambda$ scans a wide  
range of values, which change randomly, both increasing and decreasing relative to the exterior.
This is a quantum random walk \cite{Kaloper:2022oqv,Kaloper:2022utc}, and the 
variation of $\Lambda$ essentially defines a toy model of 
eternal inflation \cite{Linde:2015edk}. On average, $\Lambda$ decreases inside a sequence membranes. 

To make sure that the range of $\Lambda$
comes arbitrarily close to zero without fine tuning, we invoke a system of 
two $4$-forms that are degenerate on shell with 
$\Lambda$. We take their membrane charges such that their ratio is 
an irrational number. In this case the spectrum of values of $\Lambda$ is a 
very fine discretuum \cite{Banks:1991mb}. As noted, the general dynamical drift is to decrease $\Lambda$. 
With the specific choice of membrane charge to tension
ratio, 
 \be
 |q_j| = \frac{2 \mpl^4 |{\cal Q}_j|}{3{\cal T}^2_j} < 1 \, , 
 \label{chargeratio}
 \ee
 the only processes which are kinematically allowed for the discharge  
 favor the terminal value 
$\Lambda \rightarrow 0^+$. Near it, the discharges automatically damp down because the discharge rate 
$\Gamma \simeq \exp({-\frac{24\pi^2 \mpl^4}{\Lambda}})$ 
has an essential singularity at $\Lambda \rightarrow 0^+$ 
\cite{Coleman:1980aw}. This solves the `classic' cosmological constant problem 
\cite{Zeldovich:1967gd,Wilczek:1983as,Weinberg:1987dv} 
(sometimes also called the `old' cosmological constant problem).

The theory avoids the empty universe problem of
\cite{Abbott:1984qf}, and admits inflation. The empty universe problem of \cite{Abbott:1984qf} arose
because the dynamical adjustment mechanism devised there employed a scalar field in permanent slow
roll on an almost-linear potential, with obstacles to classical motion appearing only
near $\Lambda \rightarrow 0$. To get there, the field had to dominate the cosmic contents eternally,
supporting inflation all the way to almost Minkowski space. This meant, there was no reheating and
no matter was produced in the terminal geometry. 

Here we avoid this problem since the relaxation of $\Lambda$ 
involves large successive jumps, which come from the charges ${\cal Q}_j$ being 
large, and the tiny terminal $\Lambda$ comes from
the irrational ratio of charges. As a result, the cosmological constant does not always dominate, but just 
sometimes \cite{Bousso:2000xa}. The relaxation process is not a slow roll, but a quantum random walk. Thus
it is perfectly possible that the universe selects the terminal vacuum well before the end of the last stage
of inflation, which solves the usual cosmological problems and reheats the universe. 
Therefore a `normal' late cosmology can be embedded in our
framework. The cosmological constant problem reduces to ``{\it Why now?"}, whose answers 
might involve late time physics. We leave the questions on how to embed inflation and model late acceleration 
for later. 

Our action is a simplified version of the theory 
in \cite{Kaloper:2022oqv,Kaloper:2022utc}, given explicitly in terms of the dual magnetic 
variables, 
\ba
S &=& \int d^4x \Bigl\{\sqrt{g} \Bigl(\frac{\mpl^2}{2} R 
- \mpl^2 \bigl(\lambda + \hat \lambda \bigr) 
- {\cal L}_{\tt QFT} \Bigr)- 
\frac{\lambda}{3} {\epsilon^{\mu\nu\lambda\sigma}} \partial_\mu {\cal A}_{\nu\lambda\sigma}
-  \frac{\hat \lambda}{3} {\epsilon^{\mu\nu\lambda\sigma}} 
\partial_\mu \hat {\cal A}_{\nu\lambda\sigma} \Bigr\}  \nonumber \\
&& + \,\, S_{\tt boundary} - {\cal T}_A \int d^3 \xi \sqrt{\gamma}_A - {\cal Q}_A \int {\cal A}  
- {\cal T}_{\hat A} \int d^3 \xi \sqrt{\gamma}_{\hat A} - {\cal Q}_{\hat A} \int \hat {\cal A}  \, .
\label{actionnewmemd} 
\ea
$S_{\tt boundary}$ is a generalization of the 
Israel-Gibbons-Hawking boundary action \cite{israel,gibbhawk,gibbhawkcosm} to include 
boundary terms for the two gauge sectors \cite{Duff:1989ah,Duncan:1989ug},
\be
S_{\tt boundary} = \int d^3 \xi \Bigr( [\frac{\lambda}{3} {\epsilon^{\alpha\beta\gamma}} {\cal A}_{\alpha\beta\gamma}]
+ [\frac{\hat \lambda}{3} {\epsilon^{\alpha\beta\gamma}} 
\hat {\cal A}_{\alpha\beta\gamma}] \Bigr)  - \int d^3 \xi \sqrt{\gamma} \mpl^2[ K ]\, ,
\label{boundact}
\ee
and $[...]$ is the jump across a membrane. The charge terms are 
\be
\int {\cal A} = \frac16 \int d^3 \xi {\cal A}_{\mu\nu\lambda} \frac{\p x^\mu}{\p \xi^\alpha} \frac{\p x^\nu}{\p \xi^\beta} 
\frac{\p x^\lambda}{\p \xi^\gamma} \epsilon^{\alpha\beta\gamma} \, ,
\ee
and likewise for $\hat {\cal A}$. Here ${\cal T}_i, {\cal Q}_i$ are the membrane tension and charge, while 
$\xi^\alpha$ are membrane coordinates and embedding maps  are 
$x^\mu = x^\mu(\xi^\alpha)$. The winding direction of these maps sets 
the sign of the charge. We will take ${\cal T}_i >0$ to exclude negative local 
energy. Note that the first line of Eq. (\ref{actionnewmemd}) is a
minute generalization of unimodular formulation of GR \cite{Einstein:1919gv,Anderson:1971pn,Aurilia:1980xj,Duff:1980qv,Buchmuller:1988wx,Buchmuller:1988yn,Henneaux:1989zc,Ng:1990xz,Buchmuller:2022msj};
our full action (\ref{actionnewmemd}) generalizes it further by adding membranes.

Quantum mechanically, the membranes can nucleate in background fields \cite{Brown:1987dd,Brown:1988kg}. 
Hence these processes change the distribution of sources 
and the evolution of bubble interiors. The classical superselection sectors  
all mix up. This induces the evolution in the space of geometries due to 
the variation of $\lambda, \hat \lambda$. We remind that
here $\mps$ is fixed. We also remind that the charges ${\cal Q}_A$ and $\hat {\cal Q}_A$ have an irrational ratio,
\be
\frac{{\cal Q}_{\hat A}}{{\cal Q}_A} = \omega \in \Bigl\{{\rm Irrational ~numbers} \Bigr\}  \, ,
\label{irrational} 
\ee
as in the irrational axion proposal \cite{Banks:1991mb}. 

Further note, that (\ref{actionnewmemd}) depends on the flux variables $\lambda, \hat \lambda$ {\it linearly}, as opposed quadratically 
(the latter dependence being the common case as in \cite{Brown:1987dd,Brown:1988kg} and followup work). This has crucial importance 
for ceasing decay of the cosmological term when it approaches zero. We note that even if the higher order corrections are included,
since their weighing is by $\Mpl$, the linear terms remain dominant for sub-Planckian fluxes and the same behavior as we will uncover
below remains. Finally, the higher-order corrections could come in with different coefficients for the two flux sectors. This would induce
mutually irrational variation of fluxes even if the actual charge ratio were a rational number. We will keep this in mind as a possible
explanation of the origin of our framework, and for simplicity's sake retain only the linear fluxes and irrational ratios below. 

Now we can turn to studying the effects of quantum membrane discharge in the semiclassical limit. This means, we consider the dynamics
described by the action (\ref{actionnewmemd}) in Euclidean time, which controls the nucleation processes and their rates 
\cite{Coleman:1977py,Callan:1977pt,Coleman:1980aw}.  As explained in \cite{Kaloper:2022oqv,Kaloper:2022utc}, 
we Wick-rotate the action using $t = - i x^0_E$, defining the Euclidean action by $i S = - S_E$ and restricting to 
locally maximally symmetric backgrounds. Those are the configurations with local $O(4)$ symmetry
which dominate in semiclassical limit since they have minimal Euclidean action \cite{Coleman:1977py,Callan:1977pt,Coleman:1980aw}. 
Ergo, we set $\langle {\cal L}^E_{\tt QFT} \rangle = \Lambda_{\tt QFT}$, 
with $\Lambda_{\tt QFT}$ the matter sector vacuum energy to an arbitrary loop order. The resulting Euclidean action is 
\ba
S_E&=&\int d^4x_E \Bigl\{\sqrt{g} \Bigl(-\frac{\mpl^2}{2} R_E + \mpl^2 (\lambda + \hat \lambda) 
+ \Lambda_{\tt QFT} \Bigr)- \frac{\lambda}{3} {\epsilon^{\mu\nu\lambda\sigma}_E} \partial_\mu {\cal A}^E_{\nu\lambda\sigma}
- \frac{\hat \lambda}{3} {\epsilon^{\mu\nu\lambda\sigma}_E} \partial_\mu \hat {\cal A}^E_{\nu\lambda\sigma}\Bigr\} \nonumber \\
\label{actionnewmemeu}
&& +~S_{\tt boundary} + {\cal T}_A \int d^3 \xi_E \sqrt{\gamma}_A - \frac{{\cal Q}_A}{6} \int d^3 \xi_E \, {\cal A}^E_{\mu\nu\lambda} \, 
\frac{\p x^\mu}{\p \xi^\alpha} \frac{\p x^\nu}{\p \xi^\beta} 
\frac{\p x^\lambda}{\p \xi^\gamma} \epsilon_E^{\alpha\beta\gamma} \\
&& ~~~~~~+~  {\cal T}_{\hat A} \int d^3 \xi_E \sqrt{\gamma}_{\hat A}  - \frac{{\cal Q}_{\hat A}}{6} \int d^3 \xi_E \, 
\hat {\cal A}^E_{\mu\nu\lambda} \, \frac{\p x^\mu}{\p \xi^\alpha} \frac{\p x^\nu}{\p \xi^\beta} 
\frac{\p x^\lambda}{\p \xi^\gamma} \epsilon_E^{\alpha\beta\gamma} \, . \nonumber
\ea
From the QFT/gravity couplings, it follows that $\Lambda_{\tt QFT} = {\cal M}_{\tt UV}^4 
+ \ldots \equiv \mps  {\cal H}^2_{\tt QFT}$, where
${\cal M}_{\tt UV}^4$ is the QFT UV cutoff and ellipsis denote subleading terms \cite{Englert:1975wj,Arkani-Hamed:2000hpr}. 
Thus we can collect all the terms as $\Lambda = \mpl^2 \bigl({\cal H}^2_{\tt QFT} + \lambda + \hat \lambda \bigr) = \mpl^2 \lambda_{\tt eff}$.
From here on we drop the index ${E}$. 

The membranes serve as boundaries of regions with $\Lambda_{out/in}$, 
(where {\it out/in} denote exterior (parent) and interior (offspring) of 
the membranes, respectively)). Both {\it in} and {\it out} have the metrics of the form $ds^2_E =  dr^2 + a^2(r) \, d\Omega_3$ 
where $d\Omega_3$ is metric on a unit $S^3$ and $a$ solves $\bigl(\frac{a'}{a}\bigr)^2 - \frac{1}{a^2} 
= - \Lambda/3 \mpl^2$, and the prime is the $r$-derivative. On a membrane, the
jump of the metric is controlled by the boundary conditions, which impose that $a, {\cal A}, \hat {\cal A}$ are 
continuous, and the discontinuities are \cite{Kaloper:2022oqv,Kaloper:2022utc}
\be
\lambda_{out} - \lambda_{in}  = \frac{{\cal Q}_A}{2} \, , ~~~~ \hat \lambda_{out} - \hat \lambda_{in}  = \frac{{\cal Q}_{\hat A}}{2} \, ,
~~~~ \mps \Bigl(\frac{a_{out}'}{a} - \frac{a_{in}'}{a} \Bigr)
 = -\frac{{\cal T}_A + {\cal T}_{\hat A}}{2} \, .
 \label{metricjc}
\ee 
We compactified notation here by writing the junction 
conditions as if $A$ and $\cal A$ membranes were nucleated simultaneously. Generally,
in these equations one takes either $A$ or $\hat A$ terms.

Now to compute the membrane nucleation rates, 
$\Gamma \sim e^{-S_{bounce}}$ \cite{Coleman:1977py,Callan:1977pt,Coleman:1980aw}, we need
to construct the Euclidean instantons -- i.e. a section of 
the parent and an offspring geometry glued together along a membrane as an interface. 
The bounce action is defined by 
$S({\tt bounce}) = S({\tt instanton}) - S({\tt parent})$. The instanton taxonomy was proffered in
\cite{Brown:1987dd,Brown:1988kg} for the theories with $4$-form fluxes screening 
the cosmological constant 
\cite{Brown:1987dd,Brown:1988kg,Duncan:1989ug,Bousso:2000xa,Feng:2000if}. A comprehensive 
related analysis for the case with linear flux dependence was given in  \cite{Kaloper:2022oqv,Kaloper:2022utc}.
Here we will merely use those results. 

\begin{figure}[bth]
    \centering
    \includegraphics[width=7.5cm]{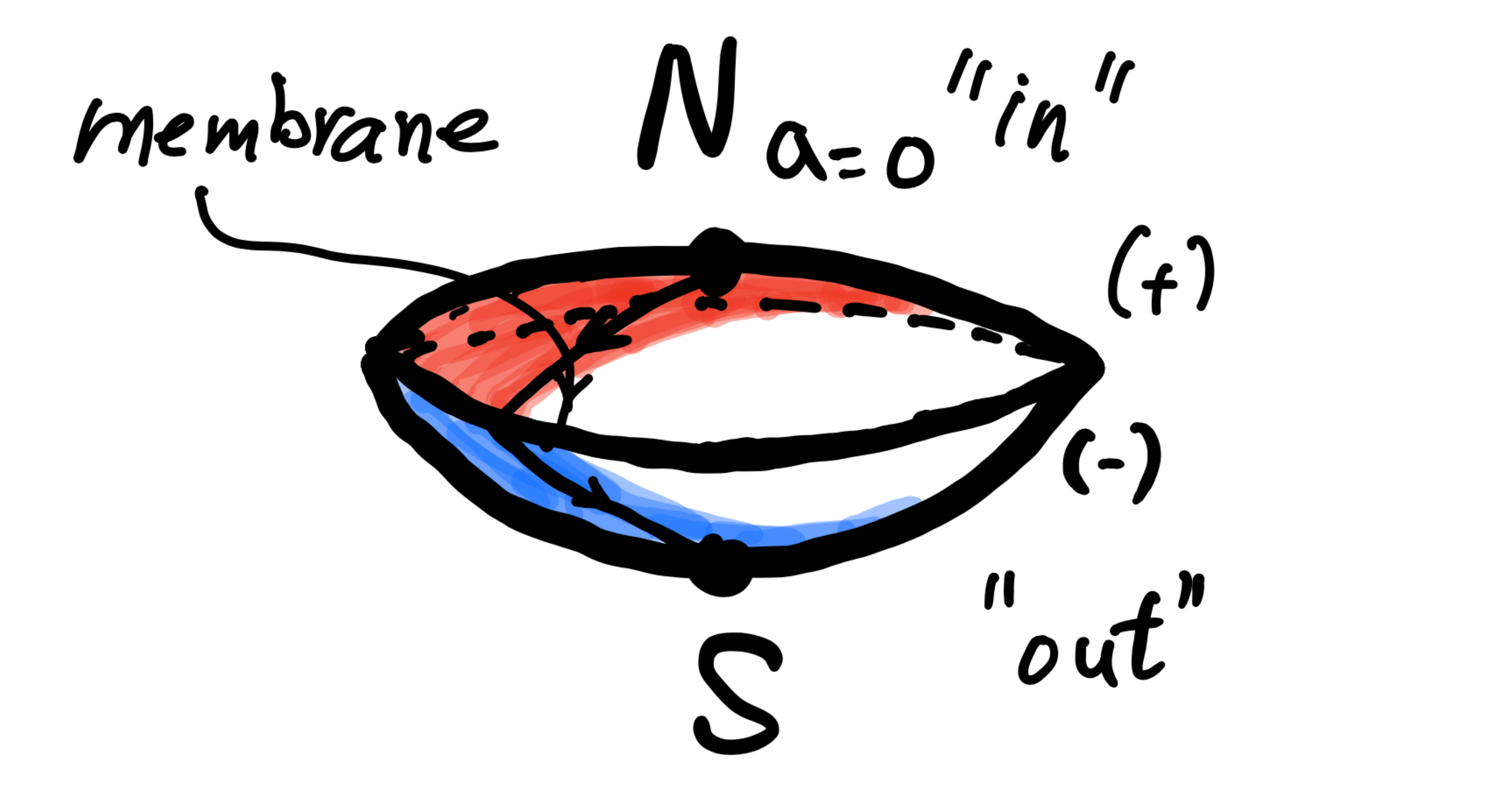}
    \caption{A $q_j<1$ instanton comprised of two sections of $S^4$.}
    \label{fig1}
\end{figure}

A key result in  \cite{Kaloper:2022oqv,Kaloper:2022utc} is that when our 
Eq. (\ref{chargeratio}) holds, $|q_j|<1$, the only transitions which are
allowed are one channel where $dS \rightarrow dS$ and one where $dS \rightarrow AdS$. 
The reason can be gleaned as follows. Combining the bulk equation for $a'$ and the 
junction conditions on the membrane
for either $A$ or $\hat A$ membrane processes, we 
find after a straightforward computation \cite{Kaloper:2022oqv,Kaloper:2022utc}
\be
\zeta_{out} \sqrt{ 1-  \frac{\Lambda_{out} a^2}{3 \mps}} 
= - \frac{{\cal T}_j}{4\mps}\Bigl(1 -  q_j \Bigr)\, a \, , ~~~~~~~~ 
\zeta_{in} \sqrt{1-  \frac{\Lambda_{in} a^2}{3 \mps}} 
= \frac{{\cal T}_j }{4\mps}\Bigl(1 +q_j \Bigr) \, a \, . 
\label{diffroots}
\ee
Here $\zeta_i = \pm$ designates two possible branches of the square root of $\bigl(\frac{a'}{a}\bigr)^2 - \frac{1}{a^2} 
= - \Lambda/3 \mpl^2$, and fixing it is required to solve the 
junction conditions (\ref{metricjc}). It is now straightforward to 
check that the Eqs. (\ref{diffroots}) allow only 
$(\zeta_{out}, \zeta_{in}) = (-,+)$ when both $|q_j| < 1$ and both membrane tensions 
are positive. Further inspection shows that the parent geometry 
must be $dS$ whereas the offspring can be either $dS$ or $AdS$.
So it turns out that the conditions $|q_j|<1$ and ${\cal T}_j > 0$ 
are extremely restrictive: only two nucleation channels are
available to both membrane systems. The other channels are either kinematically completely prohibited, or are 
suppressed by infinite bounce action \cite{Kaloper:2022oqv,Kaloper:2022utc}. Basically, what happens is that  
due to the linear dependence of the junction conditions (\ref{diffroots}) on charges -- instead of quadratic -- other 
instantons are forbidden when $|q_j|<1$ \cite{Kaloper:2022oqv,Kaloper:2022utc}. 
This effect is actually a consequence of
the gravitational vacuum stabilization found by Coleman and De Luccia 
\cite{Coleman:1980aw} which now happens for 
all values of the curvature radius due to the  linearity of (\ref{diffroots}) in ${\cal Q}_j$. 
Thus the main process of interest to us
describing $dS \rightarrow dS$ transitions is given by the instanton of Fig. \ref{fig1}. 
When $\Lambda_{out} > \Lambda_{in}$ this describes a discharge, 
and its' `time-reversal' with $\Lambda_{out} < \Lambda_{in}$ describes
an upcharge. Both are possible, but the decrease of $\Lambda$ is more likely. 

To understand the quantum discharges it is useful to solve Eqs.
(\ref{diffroots}) for $a^2$: 
\be
\frac{1}{a^2} = \frac{\Lambda_{out}}{3\mps} + \Bigl(\frac{{\cal T}_j}{4 \mps}\Bigr)^2 
\Bigr(1 -q_j\Bigr)^2 
= \frac{\Lambda_{in}}{3\mps} + \Bigl(\frac{{\cal T}_j}{4 \mps}\Bigr)^2 
\Bigr(1 + q_j \Bigr)^2 \, .
\label{radii}
\ee
This shows there are two regimes of membrane nulceations for both $A, \hat A$. If $a^2$ is comparable to 
de Sitter radii, then from Eq. (\ref{radii})  $\sim (1-\frac{\Lambda_j a^2}{3 \mps})^{1/2} \ll 1$
and so the bounce action is approximately 
\be 
S_{\tt bounce} \simeq - \frac{12\pi^2 \mpl^4 \Delta \Lambda}{\Lambda_{out} \Lambda_{in}} \, , ~~~~~~~~ 
\Delta \Lambda = \Lambda_{out} - \Lambda_{in} = \frac12 \mpl^2  {\cal Q}_j \, .
\label{fastbounce}
\ee
Since $|q_j|<1$, in this regime the discharge of the cosmological constant is fast because $S_{\tt bounce} <0$,
 as long as 
$\Lambda_{out} \gg 3 \mpl^2 \Bigl(\frac{{\cal T}_j}{4 \mpl^2}\Bigr)^2$. 
The cosmological constant decreases fast from near the Planckian scales. 
The reverse processes increasing $\Lambda$ ($\Delta \Lambda < 0$) 
have a positive action (sign-reversed (\ref{fastbounce})) and so they are 
rarer. As claimed above, the dominant trend is to decrease $\Lambda$. 

This ends when 
$\Lambda < 3 \mpl^2 \Bigl(\frac{{\cal T}_j}{4 \mpl^2}\Bigr)^2$. For smaller cosmological constants, 
the discharge nucleations proceed via production of small bubbles, with the 
bounce action \cite{Kaloper:2022utc}
\be
S_{\tt bounce}  \simeq \frac{24\pi^2 \mpl^4}{\Lambda_{out}} 
\Bigl(1- \frac{8}{3} \frac{\mpl^2  \Lambda_{out}}{ {\cal T}_j^2} \Bigr)\, ,
\label{familiars2}
\ee 
and $S_{\tt bounce} > 0$ because $\Lambda < 3 \mpl^2  \Bigl(\frac{{\cal T}_j}{4 \mpl^2}\Bigr)^2$. 
This action has a remarkable property that it \underbar{\it diverges} as $\Lambda_{out} \rightarrow 0$. 
As a result the bubbling rate $\Gamma \sim e^{-S_{bounce}}$
has an essential singularity at $\Lambda_{out} \rightarrow 0$, where the rate goes to zero. 
Hence when $|q_j| < 1$ small cosmological constants become very long lived, 
and the closer the geometry gets to a locally Minkowski  space, the more stable it becomes to discharges. If
it ends up with zero cosmological constant, further discharge stops. 

To recapitulate, we have given a theory where the cosmological constant 
is unstable to quantum-mechanical, nonperturbative, discharge 
of membranes, whose flux is degenerate with the cosmological constant 
due to covariance. The instability stops when 
$\Lambda/\mpl^4 \rightarrow 0$. This feature is a consequence of 
Coleman and De Luccia's `gravitational stabilization' of
flat space to nonperturbative instabilities, and it is operational when our Eq. (\ref{chargeratio}) holds. For the theory 
(\ref{actionnewmemd}) this holds throughout its range of validity. 
This suffices to relieve the cosmological constant problem.
Let us explain how. 

In our theory (\ref{actionnewmemd}), as noted above, the total cosmological 
constant is 
\be
\Lambda_{\tt total} = \mpl^2 \lambda_{eff} = \mps \Bigl({\cal H}_{\tt QFT}^2 + \lambda + \hat \lambda \Bigr) \, .
\label{cctotal}
\ee
Since $\lambda$ and $\hat \lambda$ change discretely, by $\Delta \lambda_j = {\cal Q}_j/2$, we have 
$\lambda_j = \lambda_{j~0} + N_j \frac{{\cal Q}_j}{2}$. For simplicity we absorb $\lambda_{j~0}$ 
into ${\cal H}_{\tt QFT}^2$. This leaves us with, using Eq. (\ref{irrational}),
\be
\Lambda_{\tt total} = \mpl^2 \Bigl({\cal H}_{\tt QFT}^2 + \frac{{\cal Q}_A}{2} \bigl( N + 
\hat N \omega \bigr) \Bigr) \, . 
\ee
Now, since we demand that $\omega$ is irrational, there exist integers $N, \hat N$ for any real number $\rho$ 
such that $N + \hat N \omega$ is arbitrarily close to $\rho$ \cite{Banks:1991mb,niven}. Therefore integers 
$N, \hat N$ exist such that 
$N + \hat N \omega$ is arbitrarily close to $- \frac{2{\cal H}_{\tt QFT}^2}{{\cal Q}_A}$. As a consequence 
there is a dense set of $\Lambda_{total}$, with values arbitrarily close to zero! In turn, this implies that 
for any initial value of $\Lambda$, there exist many sequences of discharging membranes, in any order, 
which will arrive to $N, \hat N$ at which point the cosmological 
constant is arbitrarily close to zero, and the underlying nearly
flat space is very long lived. The key reason for this is the pole of the 
bounce action, Eq. (\ref{familiars2}), which occurs for the 
$(-+)$ instantons of Fig. (\ref{fig1}), which are the only ones allowed in our case because of Eq. (\ref{chargeratio}). 
As a consequence $\Lambda \rightarrow 0^+$ is the dynamical attractor. 

This is captured by the semi-classical Euclidean partition function. 
Indeed, let's estimate  
$Z = \int \ldots {\cal DA} {\cal D} \hat {\cal A}  {\cal D} 
\lambda {\cal D} \hat \lambda  {\cal D} g  \ldots \,  e^{-{\cal S}_E}$  
by the semi-classical saddle point approximation result, 
\be
Z = \sum_{instantons} \sum_{\lambda, \hat \lambda}  e^{-{\cal S}_E(instanton)} \, ,
\label{sumZ}
\ee
where we sum over classical extrema of the action. This means, we sum 
over the Euclidean instantons with any number of membranes included. 
Since $O(4)$ invariant solutions minimize the action 
\cite{Coleman:1977py,Callan:1977pt,Coleman:1980aw}, 
in our case $Z$ should be dominated by our instantons. 

Without the explicit resummation, we can still get a feel for individual contributions. If we invert the 
bounce action, ${\cal S}({\tt instanton}) = {\cal S}({\tt bounce}) + {\cal S}({\tt parent})$, and recall 
that without offspring, the instanton action is the parent action -- i.e. the negative of the horizon area 
divided by $4G_N$, ${\cal S}({\tt parent}) = - 24 \pi^2 \frac{\mpl^4}{\Lambda_{out}}$ -- we can see that every time
a transition occurs we add a bounce action for the process to the parent action. E.g. 
consider a sequence of nested segments separated by 
membranes. By Eq. (\ref{familiars2}), a segment's contribution to 
the total action is 
${\cal S}({\tt instanton}) \la - 64 \pi^2 {\mpl^6}/{\cal T}_i^2$. A long sequence may yield  
${\cal S}({\tt instanton},n) \rightarrow - 64 \pi^2 {\mpl^6} (\frac{n_A}{ {\cal T}_A^2} 
+ \frac{n_{\hat A}}{{\cal T}_{\hat A}^2} ) \rightarrow  - 24 \pi^2 \frac{\mpl^4}{\Lambda_{\tt terminal}}$.
Thus the total instanton action for a large sequence of nucleations of both types of membranes will be bounded by 
a number of contributions, which implies the total cannot exceed $- 24 \pi^2 \frac{\mpl^4}{\Lambda_{\tt terminal}}$. 
This happens since nucleations can go on until $\Lambda_{\tt terminal} \rightarrow 0^+$. Beyond it, 
the processes that involve a $dS \rightarrow AdS$ jump are allowed, but there can only be one such jump if 
it ever occurs: once a sequence ends up in $AdS$, membranes
will not nucleate any further because $|q_j| <1$. The dynamics 
stops, and $AdS$ is a terminal sink \cite{Linde:2006nw}. 

As a consequence the semiclassical partition function 
(\ref{sumZ}), $Z \sim \sum e^{24 \pi^2 \frac{\mpl^4}{\Lambda} + \ldots}$, 
is dominated by configurations for which $\Lambda \rightarrow 0$, 
given that discharges cease in the Minkowski limit. 
Thus our mechanism super-exponentially 
favors\footnote{For earlier arguments see \cite{Hawking:1981gd,Baum:1983iwr,Hawking:1984hk}.}
\be
\frac{\Lambda}{\mpl^4} \rightarrow 0^+ \, . 
\ee
de Sitter is unstable, and quantum mechanics + GR 
dynamically relax $\Lambda$ to zero. The final (near) Minkowski space is extremely long lived.


 \begin{figure}[bth]
    \centering
    \includegraphics[width=6cm]{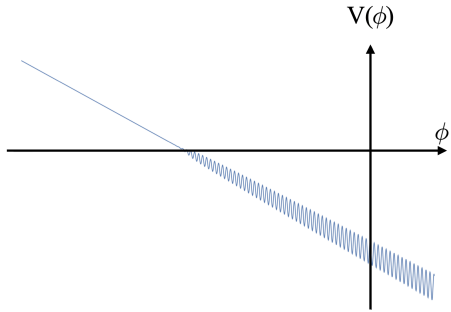}
    \caption{A schematic of Abbott's adjustment mechanism.}
    \label{figa}
\end{figure}

Our mechanism shares some features with the very insightful paper 
by Abbott \cite{Abbott:1984qf}, who designed a field-theoretic
adjustment mechanism using a scalar with a potential given by a linear 
term and a strong-coupling induced harmonic modulation.
Due to the universality of gravity, the potential and the cosmological constant 
were degenerate, and thanks to an approximate
shift pseudo-symmetry the scalar was screening away the cosmological term. However since the scalar evolution
was purely classical slow roll, to adjust $\Lambda$ to zero the scalar 
had to dominate the stress energy tensor forever - or at least 
until the cosmological constant were nearly zero. Only then did the 
harmonic modulations kick in and arrest the scalar.
As a result, the universe was inflating forever, and no reheating was possible until the Hubble parameter relaxed to 
$\simeq 10^{-34} {\rm eV}$ - which means, the universe never reheated. 
This made the victory pyrrhic, and led to the 
empty universe problem. 

In our case the empty universe problem of \cite{Abbott:1984qf} is averted since the relaxation of $\Lambda$ 
involves large successive jumps, because the charges ${\cal Q}_j$ are large. 
Every time a membrane is nucleated, the cosmological constant jumps by a 
large step $\propto \mps {\cal Q}_j$. Since the charges are large, being only subject to the constraint
(\ref{chargeratio}), the cosmological constant term can change down to its future terminal value faster than the
age of the universe now. The tiny terminal $\Lambda$ comes not from small individual charges, but from their
misalignment, which arises due to the irrational ratio of charges in Eq. (\ref{irrational}). The tiny terminal value is not
necessary, but it is super-exponentially favored from all other values 
by evolution -- because the small values are most 
long-lived. As a result, the cosmological constant does not always dominate, but just 
early on \cite{Bousso:2000xa}. 

A related point is that for exactly this same reason, inflation may also be embedded 
in this framework. Inflation should happen as the membrane evolutions stop, and the reason may be that by the time
the universe starts to inflate in slow roll, the remaining relaxation of the net vacuum energy by slow roll is faster
than that by quantum membrane nucleation. If after the universe 
exits slow roll inflation, there is a net large cosmological
constant, that will not be a terminal state and the evolution will continue. The chances for finding the right history are
further assisted by the possibility that even an empty 
universe could `restart' itself by a rare quantum jump. If it increases 
the cosmological constant, in subsequent evolution an inflationary stage could be found \cite{Garriga:2000cv}. Thus 
even a `rare' inflation can be found eventually \cite{Carroll:2004pn}. Therefore it seems that a `normal' cosmology 
can be embedded in our framework.  Precisely what the likelihood is that inflation does occur after the membrane
nucleations end is a question beyond this work. For now we must take solace from the fact that this
is not excluded. Having a cosmological constant to be, most likely, tiny 
without deploying anthropics certainly helps ease the pain. 

\vskip.7cm

{\bf Acknowledgments}: 
We would like to thank W. Buchmuller, G. D'Amico and A. Hebecker 
for valuable comments and discussions. 
NK is supported in part by the DOE Grant DE-SC0009999. AW is 
supported by the ERC Consolidator Grant STRINGFLATION 
under the HORIZON 2020 grant agreement no. 647995.

\end{document}